# The gas flow pattern through small size Resistive Plate Chambers with 2mm gap

Yousef Pezeshkian*[1], Amir Kiyoumarsioskouei[2], Majid Ahmadpouri[1], Ghasem Ghorbani[1]

[1]Department of Physics, Sahand University of Technology, Tabriz, East Azerbaijan, 51335-1996, Iran

[2]Faculty of Mechanical Engineering, Sahand University of Technology, Tabriz, East Azerbaijan, 51335-1996, Iran

## Abstract

A prototype of a single-gap glass Resistive Plate Chamber (RPC) is constructed by the authors. To find the considerations required for better operation of the detector's gas system, we have simulated the flow of the Ar gas through the detector by using computational fluid dynamic methods. Simulations show that the pressure inside the chamber linearly depends on the gas flow rate and the chamber's output hose length. The simulation results were compatible with experiments. We have found that the pressure-driven speed of the gas molecules is two orders of magnitude larger in the inlet and outlet regions than the blocked corners of a $14 \times 14$ cm$^2$ chamber, and most likely seems to be higher for larger detectors and different geometries.

## 1 Introduction

Resistive Plate Chambers (RPCs) are widely used gaseous detectors in experimental particle physics with growing applications in medical and industrial imaging [1]. RPCs are attractive not only because of their great achievable time and position resolutions and hit rate capabilities but also for their simplicity. Its structure is similar to a capacitor with dielectrics whose electrodes are connected to a high voltage (HV) power supply, and a specific gas mixture is flowing inside the gap (between plates of the capacitor). Suitable adjustment of the HV makes this capacitor a particle detector, where a passage of a charged particle may ionize few particles of gas molecules and initiate an electron avalanche drifting toward the anode.

Gaseous detectors, like RPCs, demand gas mixing systems besides the main components of the detectors [2], [3]. Further considerations regarding the gas flow patterns and their effects on the detector's operation are needed. Gas leakage from RPCs is essential, especially when an experiment requires the operation of detectors for a long time.

---

* Corresponding author.
E-mail address: pezeshkian@sut.ac.ir

High Global Warming Potential (GWP) of C2H2F4, the main component of RPC's gas mixture, recently caused further attention to the gas system of RPCs. In addition to the search for new gas mixtures with no hydrofluorocarbon (HFC) ingredients [4], [5], several attempts have been made to reduce the gas flow rate [6], [7] and even to construct and test detectors performance when there is no flow of the gas, i.e., sealed RPCs [8], [9]. Concerns of the RPC community encouraged us to simulate the effects of changing gas flows on the performance of RPCs.

Vanheule, in his thesis, simulated the flow of gas through the RPC's chamber by using the Gerris package [11]. Here, we have simulated the flow of the Ar gas through the hoses and the chamber itself with a commercial ANSYS-Fluent package [10]. Since our facility in the laboratory (at the Sahand University of Technology) was not accurate for low flow rates and to compare the simulation results with experiments, we did the simulation at relatively higher flow rates.

We outlined the paper as follows: In section 2, we introduced RPCs constructed at our laboratory. Sections 3 and 4 review the simulation procedure results, respectively.

## 2  RPCs constructed at SUT, and the gas leakage test

Our primary consideration for constructing RPCs in SUT was to use materials available in the near market at a low price, besides keeping the detector's performance in good operational conditions. In the prototype of this study, we have used commercial glass sheets with dimensions of $17 \times 17$ cm$^2$ to construct an RPC with an internal volume of $14 \times 14 \times 0.2$ cm$^3$. For simplicity, we have designed a single $10 \times 10$ cm$^2$ pad to read out the signals. High Voltage (HV) electrodes are 2 cm wider than the pads (1 cm from each side) to avoid electric field non-uniformities at the edges of electrodes. We have also utilized the medical infusion hoses for the gas nozzles (Figure 1) and Ar as the working gas in this study.

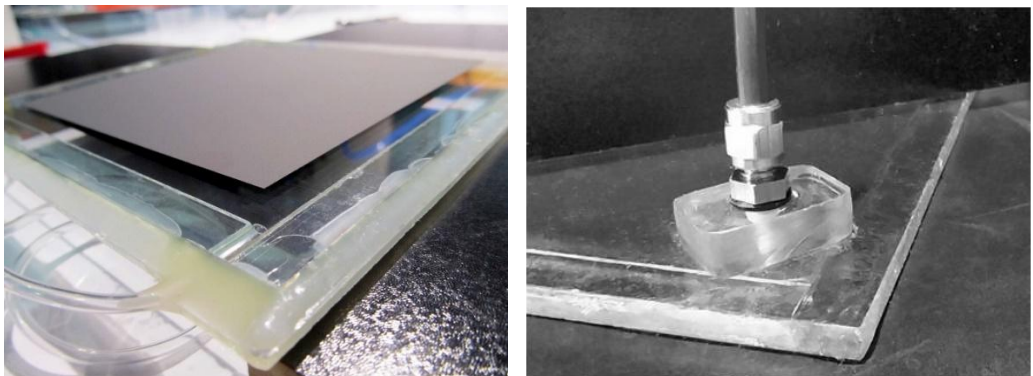

*Figure 1 Connection of the infusion hoses from the side of the chamber (left); the top nozzle design (right).*

In addition to constructing a prototype in which the gas nozzles are located at the sides of the RPC chamber (Figure 1 - Left), we build another prototype with nozzles at the top surface of the chamber (Figure 1 - Right) to compare it with our primary detector. The side nozzles are easier to handle when we have to pack several detectors close together. Here, we wanted to evaluate its effect on detector's performance.

The gas leakage is negligible when we glue detectors carefully. In small time scales (about 70 minutes), we did not observe the gas leakage from the detector. Figure 2 shows the change of the pressure difference between inside and outside of the detector by time. A commercially available sensor, "BOSCH BMP280", records the pressure frequently.

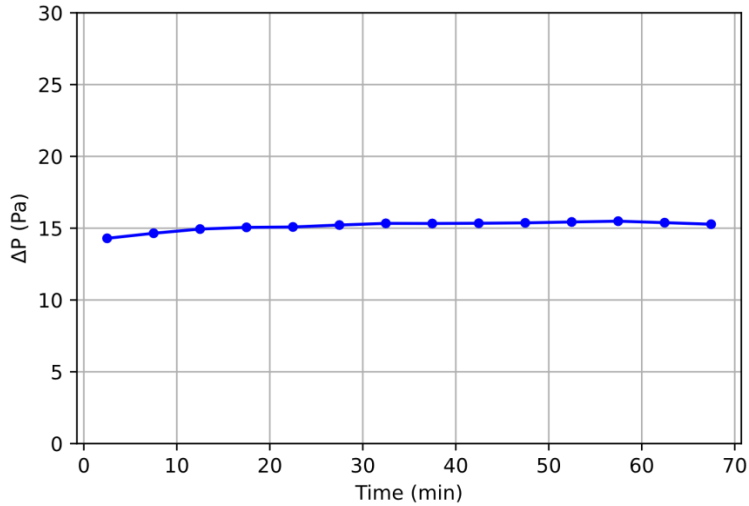

*Figure 2 Pressure variation inside the chamber is recorded continuously for 70 minutes*

## 3 Gas flow simulation

We performed a 3D CFD simulation to obtain characteristics of the gas mixture flow regime. In the simulation, we assumed that the chamber is initially filled with nitrogen gas. During the simulation time, argon replaces nitrogen. The CFD model considers a non-stationary laminar compressible transient flow. In order to avoid excessive simulation errors due to the sharp pressure gradients in cells near corner boundaries, an inflated mesh discretization was **performed** in the sharp edge situations. The aspect ratio of the quad-cells, the grid size, and the time step size were chosen fine enough to achieve mesh independence and ensure acceptable errors in flow simulations. In this study, both transient state simulation and steady-state simulation have been utilized. Transient-state simulation is used to obtain the preparation time and other unsteady phenomena, where the steady-state simulation is used to obtain the velocity distribution and pressure distribution for very long times after the start of the simulation. The resultant mesh has amounts between 40-200 thousand elements dependent on the size and domain geometry.

In the solving model, the first-order, pressure-based, implicit solver was used to solve continuity and momentum equations that lead to the pressure and velocity fields.

In this paper, we want to address several questions about the flow of the gas in RPCs. The first question is that how long the gas replacement takes? Moreover, how does this time vary with flow rate or gas pressure inside the chamber?

Keeping in mind that the chamber has four holes, the gas comes in from one hole and goes out from another, and two corners are blocked (Figure 3). We know that the speed of the gas molecules is not homogeneous inside the chamber. The second question is whether the variation of the speed of the gas molecules at different positions of the chamber is negligible or not? We want to compare this speed with the drift speed of Ar ions during the operation of the RPC.

For toxic and flammable gas mixtures, we have to guide the exhausted gas outside the laboratory. Therefore, using long hoses at the outlet of the chambers seems unavoidable. For a constant flow rate, the gas pressure inside the chamber depends on the length of the hose. Obtaining a general pattern that displays the chamber's pressure as a function of the hose's length is a crucial question in RPC design and fabrication.

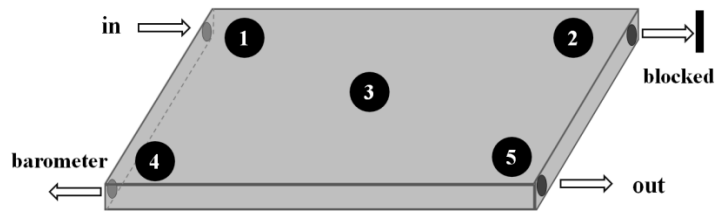

*Figure 3, Setup of experiment and simulation. Numbers show the positions in which gas parameters are obtained by simulation (see the text)*

## 4 Simulation results and discussion

### 4.1 The Preparation Time

We need to have a variable that indicates the time required to have a uniform distribution of the working gas (new gas) inside the chamber. Considering the labels of Figure 3, we defined the **preparation time** as the time needed for the average mole fraction of the Ar gas in regions 2, 4, and 5 (the critical corners of the chamber) to reach 0.99, while the mole fraction of Ar was zero at the beginning. The preparation time mainly depends on the chamber size and gas flow rate.

For the gas flow rate of 200 mL/min and the output hose length of 10 cm, Table 1 shows the simulation results for the preparation time and the gas molecule speeds. Since the gas tends to travel the shortest route, positions inside the chamber labeled by numbers 1, 3, and 5 in Figure 3 will be replaced with new gas sooner than the rest. Comparison of the two columns shows that the preparation time does not depend significantly on the inlet (outlet) orientations.

*Table 1 Gas molecules speed at five positions of the chamber specified in Figure 3*

| Gas inlet position | | side | top |
|---|---|---|---|
| Preparation time (s) | | 32.36±0.5 | 31.635±0.5 |
| Gas molecules speed At specified positions (m/s) Figure 3 | 1  in            (15mm, 15mm) | 0.237±0.001 | 0.638±0.001 |
| | 2  blocked    (125mm, 15mm) | 0.005±0.001 | 0.0025±0.0005 |
| | 3  middle     (70mm, 70mm) | 0.017±0.001 | 0.017±0.001 |
| | 4  barometer (15mm, 125mm) | 0.002±0.001 | 0.0025±0.0005 |
| | 5  out          (125mm, 125mm) | 0.052±0.001 | 0.23±0.001 |

The simulation shows that the time needed to fill regions 1, 3, and 5 is less than 2s as the mass transfer mechanism in these regions is mostly convection, but the diffusion mechanism which is dominant in regions 2 and 5 (as the speed of flow is not significant) causes the increase of the preparation time up to 32s.

To have an intuition of the gas exchange, we plotted the mole fraction of Ar in the chamber as a function of time. According to figure 4, the change proceeds faster at first and slows down when we get closer to the state of saturation.

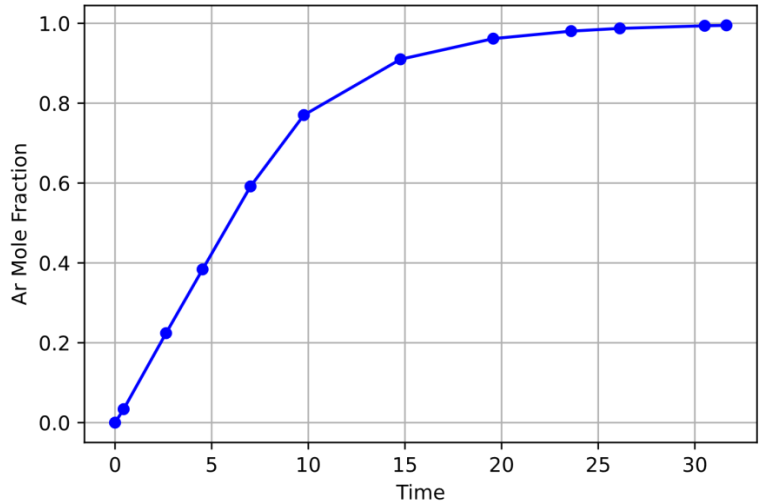

*Figure 4  Time variation of the mole fraction of argon gas inside the RPC chamber*

The pressure-driven speed of the gas molecules varies from 0.2 m/s at the entrance to 0.002 m/s at blocked corners of the chamber. The maximum pressure-driven speed of the molecules is comparable with the drift speed of Ar ions in the applied electric field. Since the electrons are responsible for signals in the RPCs, variation in the speed of ions is irrelevant in the performance of these detectors.

Variation in the speed of the gas molecules highly depends on the detector's dimensions. For a chamber with a narrow and long design (similar to a tube), speed would be approximately the same everywhere in the chamber. While, for flat chambers with equal width and length, the speed was maximum near the nozzles and minimum in the opposite blocked corners of the chamber. Sharper the blocked corners, lower the speed of the gas molecules.

In our square shape detector, the pressure-driven speed of the gas molecules varies from 0.2 m/s at the entrance to 0.002 m/s at the blocked corners of the chamber. The speed of the gas molecules depends on the size and geometry of a detector. According to Bernoulli's equation, the preparation time would be higher for larger detectors and geometries with sharper corners.

**4.2 The pressure, the flow rate, and the Output hose length**

For six flow rates from 50 to 500 ml/min, we found the pressure of the chamber by simulation (Table 2). To validate the simulation results, we measured the pressure for the same flow rates of the simulation in the experiment. Adjustment of the flow meter in the experiment was difficult, and despite our efforts, there was more than 10% inaccuracy. The difference between values obtained for the pressure in the experiment and simulation is less than 15%. By considering the operational errors and gas leakage from the chamber in the experiment, we can conclude that the results are compatible.

*Table 2- Values obtained by simulation and experiment for pressure of chambers as a function of Flow rates*

| **Flow rate (mL/min)** | **50** | **100** | **200** | **300** | **400** | **500** |
|---|---|---|---|---|---|---|
| **Pressure (sim.) (Pa)** | 13.6±0.4 | 27.8±0.7 | 58.5±1.5 | 93.5±2.4 | 131.8±3.4 | 172.0±4.4 |
| *Pressure (exp.) (Pa)* | *27.48±0.88* | *37.94±0.44* | *58.51±0.65* | *108.18±0.85* | *140.22±1.24* | *193.95±0.60* |

In the pressure range usually used for RPCs (i.e., below 300Pa or 3mbar gauge pressure), we can fit a linear function to the simulation's data (Figure 5). The pressure increases linearly by the flow rate. In other words, the flow rate is a good indicator of the pressure inside the chamber.

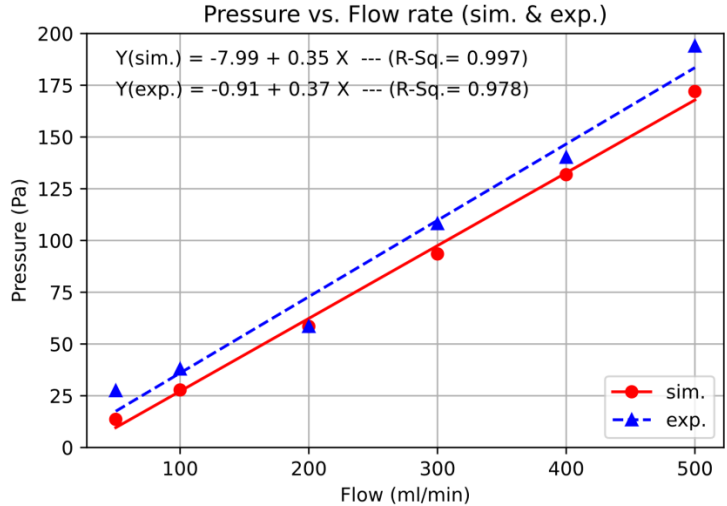

*Figure 5 Pressure inside the chamber as a function of flow rate for simulation (left) and experiment (right).*

Since the hose length of the chambers affects the pressure of the detector, and the hose length may be very long in some experiments to guide the exhausted gas outside of the laboratory, we performed another simulation to understand their relation. Figure 6 shows that the pressure will increase linearly by increasing the hose length if we keep the flow rate unchanged (300 mL/min). Increasing the hose length causes more resistance on the downstream side of fluid flow. Therefore, more pressure is needed to establish a constant flow rate.

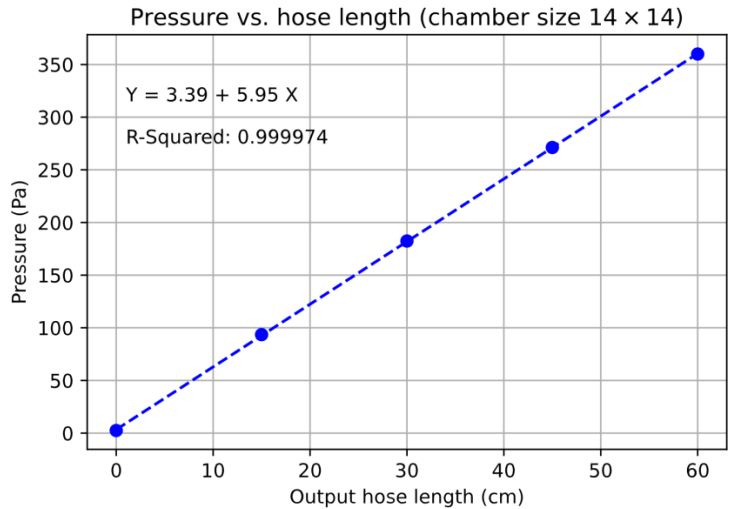

*Figure 6 Pressure inside the chamber as a function of output hose length*

## 5 Conclusions

RPCs can work with low gas flows unless we use them in high hit rate environments. Measurement of the flow rate for the low Reynolds numbers, needs more accurate flowmeters than ours; we could not reduce the flow below 50 mL/min. In the simulation, to be able to validate results, we chose the same flows as experiments. In the flow rates between 50 to 500 ml/min, there was a linear relationship between flow rate and pressure and between pressure and hose length. We may extrapolate this linearity for lower flow rates. A more accurate conclusion needs further simulations.

At the beginning of the experiment, we have to replace the air in the chamber with the RPC's working gas (or gas mixture). Suppose V is the volume of the chamber. The time required to complete the gas exchange, the preparation time, is more than the time required to flow 1V of gas through the detector. The flow rate of our simulation in Table 1 is 200 ml/min or 5.1 V/minute, which means that every ~12 seconds, 1 V of gas enters the detector. In the same flow, the preparation time was about 32 s, which is ~3 times the time required to amount of gas circulation reaches 1 volume of the chamber.

The gas density and viscosity play an imposing influence in preparation time and hydrodynamic parameters of the flow field. Increasing the gas density and viscosity caused a significant decrease in gas flow rate (by the same upstream-side pressure). Also, more time is needed to fill the chamber for more dense and viscous fluids. Estimated calculations show a notable difference between the R134a and argon preparation time.

It is also clear that the fluid type has no effect on the preparation time, which means that other gases such as R134a have a similar preparation time to argon. Pressure in the chamber, on the other hand, is a function of viscosity and gas density. So the pressure would most likely be different for a different gas (or gas mixture). For R134a, the pressure will be lower due to the viscosity ratio between argon and R134a.

We have also stated that our studies may not be valid for MRPCs, which have much narrower gaps, and we are planning to repeat this study for MRPCs.